\documentclass[preprint2]{aastex}

\usepackage{epsfig} 
\usepackage{amssymb}
\usepackage{amsmath}
\usepackage{rotating}

\newcommand{\totalno}{69~}	
\newcommand{\onelineno}{15~}

\newcommand{\ps}{P\&S}
\newcommand{\U}{\textsf{U}}
\newcommand{\V}{\textsf{V}}

\newcommand{\vsini}{$v\sin i$}
\newcommand{\kms}{km$\cdot$s$^{-1}$}
\newcommand{\degg}{~deg$^2$}

\shorttitle{XO astrophysical false positives}
\shortauthors{Poleski et al.}

\begin{document}

\title{The XO planetary survey project --- \\ Astrophysical false positives}

\author{
Rados\l{}aw Poleski\altaffilmark{1},
Peter R. McCullough\altaffilmark{2},
Jeff A. Valenti\altaffilmark{2},
Christopher J. Burke\altaffilmark{2},
Pavel Machalek\altaffilmark{2,3},
Kenneth Janes\altaffilmark{4}
}

\altaffiltext{1}{University of Warsaw Observatory, Al. Ujazdowskie 4, 00-478 Warszawa, Poland}
\altaffiltext{2}{Space Telescope Science Institute, 3700 San Martin Drive, Baltimore, MD 21218, USA}
\altaffiltext{3}{Department of Physics and Astronomy, Johns Hopkins University, Baltimore, MD 21218, USA}
\altaffiltext{4}{Astronomy Department, Boston University, 725 Commonwealth Avenue, Boston, MA 02215, USA}

\email{rpoleski@astrouw.edu.pl}

\begin{abstract}
Searches for planetary transits find many astrophysical false positives as a by-product.
There are four main types analyzed in the literature: a grazing-incidence eclipsing binary 
star, an eclipsing binary star with a small radius companion star, a blend of one or more
stars with an unrelated eclipsing binary star, and a physical triple star system. 
We present a list of 69 astrophysical false positives that had
been identified as candidates of transiting planets of the on-going XO survey. This 
list may be useful in order to avoid redundant observation and characterization of 
these particular candidates independently identified by other wide-field searches for 
transiting planets. The list may be useful for those modeling 
the yield of the XO survey and surveys similar to it. Subsequent observations of some 
of the listed stars may improve mass-radius relations, especially for low-mass stars. 
From the candidates exhibiting eclipses, we report three new 
spectroscopic double-line binaries and give mass function estimations for \onelineno 
single lined spectroscopic binaries.
\end{abstract}


\section{Introduction}
\label{sec:intro}
A planetary transit indicates that the orbital inclination $i \approx 90$\arcdeg, 
so the projection factor $\sin\left(i\right)$ is near unity and thus measurements of 
the radial velocity (hereafter RV) of the star, 
which mass is known, %
reveal the true mass of the planet, $M_p$, not 
just the product $M_p \sin\left(i\right)$. Furthermore, the photometric depth of a 
transit indicates the ratio of the planetary radius to the stellar radius.

With increasing precision of the observations, planetary transits provide a wealth
of information about the physical characteristics of the planet and the star 
\cite[see e.g. ][]{charbonneau07}. Because of their importance, much effort has 
been applied to finding transiting planets and dozens have been 
reported.\footnote{{\tt http:www.inscience.ch/transits/}} The results described 
in this work originate from the XO project \citep{mccullough05}, and because the 
characteristics of that survey are similar to 
other transit surveys such 
as HAT \citep{bakos02}, WASP \citep{pollacco06}, TrES \citep{dunham04}, some of
the results may be 
helpful to avoid redundant observations for transit candidates selected by these surveys.

Planetary transit surveys produce a large number of homogeneous photometric 
measurements. \citet{paczynski00} listed many ways 
in which data collected by massive photometric surveys 
can be used for astrophysical 
research. During planetary searches many 
objects %
mimicking planetary transits are discovered: 
eclipsing binaries (EB) on grazing incidence orbits, small stars transiting larger 
stars (e.g. an M dwarf plus an F dwarf or a dwarf and a giant) or eclipsing binary 
systems with the orbital periods between 0.5~d and 10~d (typical period search criteria) 
diluted by a bright star (either physically connected or not) which 
reduces the eclipse depth to approximately 1~\%. In grazing-incidence case, the orbital 
period is two times longer than the photometric one and these systems can sometimes be 
distinguished if the odd and even eclipses have different depths, which implies 
different surface brightnesses, or the separation in time from one eclipse to the next
is bimodal, which would imply an eclipsing binary in an eccentric orbit.
It is sometimes extremely 
difficult to prove the true nature of transiting objects \cite[see e.g.][]{hoyer07}. 
\citet{creevey05}, \citet{young06} and \citet{beatty07} showed that follow up 
observations of these stellar transits can further our knowledge of the K and M dwarfs.

Transiting extrasolar planets are typically found by photometric surveys and later-on confirmed by RV observations.
More often than not, the subsequent RV 
measurements reveal not a planet but an {\it astrophysical false positive} 
\citep{brown03}. In this paper we report on some of the astrophysical false positives observed by 
the XO project \citep{mccullough05}. The structure of the paper is as follows: 
\S \ref{sec:pho} describes the photometric observations 
and %
their analysis, 
\S \ref{sec:list} describes the list of astrophysical false positives 
and compares it with other similar lists, %
\S \ref{sec:rv} discusses the radial velocity measurements  
followed by discussion of selected objects in \S \ref{sec:notes} and conclusions 
in \S \ref{sec:conc}.

\section{Photometric observations and their analysis}
\label{sec:pho}
\citet{mccullough05} described the equipment and the observation strategy associated with
the data analyzed here.
Since September 2003 two XO cameras have operated autonomously 
at Haleakala summit in Hawaii. Each camera consists of a wide-field 200~mm f/1.8 lens 
combined with a 1024$\times$1024~pixel CCD detector observing in the  0.4~\micron~to 
0.7~\micron~band-pass. The field of view is 7\fdg2$\times$7\fdg2. Each pixel is 
24~\micron~yielding an image scale of 25\farcs4 per pixel. Both cameras are attached 
to the same German-equatorial mount, operating in a drift-scan mode along the N-S strips. 
Each star is observed by both cameras every 10~minutes with 54~s exposures. On many 
nights, some stars are observed for less than 4~h so 
frequently only an ingress 
or an egress is observed. Up to 45000 stars brighter than 13.3~mag in $V$ per strip 
are analyzed for planetary transits. The standard deviation of photometric time series for stars 
brighter than 12~mag is typically less than 10~mmag. Stars brighter than 8.5~mag are 
saturated and thus not analyzed.

Here we present an analysis of seven strips each 7\fdg2~wide, covering declinations from
0\degr~to 63\degr~and centered at right ascensions 0, 4, 7.5, 8, 12, 15.5 and 
16 hours. Since we observe each strip for $\approx$4 months 
each year, for some targets only a few transits are observed which leads to possible 
period ambiguities. In present study we have used data collected during: two and a half 
seasons for 0 and 4 h RA strip, one season for 15.5 h RA strip and two seasons for the 
rest of the strips.

Two different methods were employed to correct 
photometric data for systematic errors: 
SysRem \citep{tamuz05} and one described by \cite{mccullough05}. 
Each of these methods produced one set of input data for 
planetary transit search, for which 
we used the Box-fitting Least Squares algorithm \citep[BLS; see ][]{kovacs02}, %
as modified by \citet{mccullough05}. %
For the most promising candidates we combined BLS results (depth, period and duration of the transit) with
catalog information to estimate the planetary radius 
and other ancillary facts related to the star. 

Because of the large pixel scale of the XO cameras, there is a substantial
possibility that a nearby bright star can contaminate the light from an EB, 
mimicking a planetary transit. Thus, 
for each %
candidate we calculated the fraction 
of light 
(hereafter FL75) within the XO photometric aperture ($r=75\arcsec$) from varying source using  
the 2MASS point-source catalog \citep{skrutskie06}. 
The estimate of FL75 gives possibility to better constrain undiluted transit depth:
\begin{equation}\label{equ:dmfl75}
\delta m_{FL75} = -2.5\log\left(1-\frac{1-10^{-0.4\delta m}}{\mathrm{FL75}}\right)
\end{equation}
Where $\delta m$ is transit depth found using BLS on XO data and $\delta m_{FL75}$ is 
expected undiluted depth. At the beginning of selection process it is not known which 
of the stars situated within XO
photometric aperture is varying and only lower limit on 
$\delta m_{FL75}$ can be constrained. One should note that the true transit depth 
can differ from the XO one also because BLS fits a box-shaped function to the transit 
and in most cases BLS underestimates the depth. 

One of the methods used for identification of blended double system involves centroid 
shift \citep{mccullough07,burke06}. For each candidate we compared astrometric positions 
measured during in-transit and out-of-transit observations.
In this paper we present three candidates eliminated only using centroid shift. 
Standard deviations of mean positions were $\approx0\farcs03$ in almost all cases.

Follow-up photometric observations of selected candidates are conducted by 
the XO Extended Team (hereafter ET) with higher 
angular resolution and in several band-passes. The ET is composed of amateur 
astronomers whose observing sites are dispersed widely in longitude. Their 
observations often show blended neighbors of target stars. The ET observations 
also have higher photometric precision and better timing of transits which is 
useful for improving the accuracy of the ephemerides 
and identifying the nature of transiting planet candidates.
Representative light curves from ET observations are shown in other papers \citep{burke07,burke08,mccullough06,johnskrull08,garciamelendo09}.

\section{List of astrophysical false positives}
\label{sec:list}
\begin{figure}[t]
\plotone{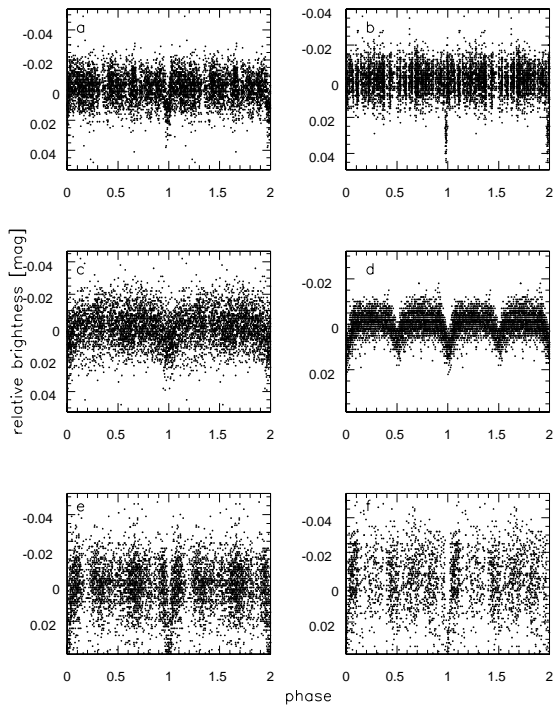}
\figcaption{Exemplary  light curves for XO astrophysical false positives. 
In all cases two cycles are shown and phase 0 corresponds to middle of the 
transit. Best ephemeris given in this paper were used. Data were calibrated 
using SysRem algorithm. Star designations: 
{\it a} -- J0401\ldots, {\it b} -- J040022\ldots, {\it c} -- J0736\ldots, {\it d} -- J0745\ldots, {\it e} -- J23503\ldots, {\it f} -- J2351\ldots.\label{fig:lcs}}
\end{figure}

Table~\ref{tab:falsepos} lists stars which 
passed automated selection criteria and showed variations similar to a planetary 
transits on the XO 
photometric %
data. We report there: 2MASS designation, the XO brightness 
($m_{XO}$), FL75, transit properties given by BLS (depth $\delta m$, duration 
$t_{dur}$, period $P$ and time of epoch $T_0$), $\delta m_{FL75}$, the type of the astrophysical 
false positive and additional remarks. Figure \ref{fig:lcs} shows exemplary light curves 
for %
astrophysical false positives.

The XO data presented here cover an area of 3539\degg, with \totalno astrophysical 
false positives. The SuperWASP fields analyzed by \citet{christian06}, 
\citet{clarkson06} and \citet{kane08} were 1507\degg, 786\degg~and 2162\degg~with 
41, 44 and 30 false positives, respectively. The common area with XO were equal to 
133\degg, 188\degg~and 361\degg, respectively. 
We identified only two objects listed in Table~\ref{tab:falsepos} in common with these lists: 
J00023\ldots and 
J1157\ldots, 
for which transits were found by SuperWASP project. These stars are described in more detail in \S \ref{sec:notes}. 
The reason for such a small number of common objects may result 
from different automated cutoff limits of calculated quantities, and 
the fact that %
the fraction 
of recovered transits does not equal unity for all periods under consideration. 
The observing window causes problems with revealing transits occurring at some periods, especially integer-day and long ones. Simulations show that 
the fields analyzed by \citet{christian06} were observed sufficiently to recover 
all transits shorter than 3~d. This is also true for half of the fields observed by 
\citet{kane08} and not true for the analysis presented by  
\citet{clarkson06}.
Another similar lists 
were presented by \citet{odonovan07}, \citet{street07} and \citet{lister07} but 
their fields did not overlap the XO fields. 

As marked in Table~\ref{tab:falsepos} 
some of astrophysical false positives have flat-bottomed transits (\U-shape) and others have long ingress and egress 
with little or no ``bottom'' (\V-shape). 
If ET observations exist they were used to assign 
\U{} or \V.
The former (\U-shaped) presumably are 
showing total eclipses and the latter (\V-shaped) are on grazing incidence orbits.
Because some of the \U-shaped 
transits may be caused by eclipsing M dwarfs (or even brown dwarfs) they are good 
targets for additional follow-up observations.

\section{Radial Velocity measurements}
\label{sec:rv}
Only a few of the XO candidates turned out to be planets 
\citep{burke07,burke08,johnskrull08,mccullough06,mccullough08}.
Table~\ref{tab:falsepos} demonstrates the importance of
the ET follow-up photometry to the XO project in order to keep the false positive rate for RV measurements low,
as spectroscopic verification requires precious time on large telescopes.
Scheduling RV measurements is straight forward because the RV is changing all the time,
whereas the planetary transit reveals itself only for a few percent of the period. Sometimes 
two RVs show a difference much larger than 1~\kms{} (planets orbiting other stars typically 
show smaller RV amplitude) and prove the companion is not a planet. The advantage of taking 
spectra is more pronounced for long-period objects or ones with periods close 
to an integer number of days, e.g. \object{OGLE-TR-111-b} whose transits were unobservable 
from northern Chile in 2007 \citep{minniti07}. 

Spectra were taken using spectrographs at: 2.7~m Harlan J. Smith Telescope (HJS) 
at McDonald Observatory, 4~m Nicholas U. Mayall Telescope at Kitt Peak National 
Observatory and 11~m Hobby-Eberly Telescope (HET) at McDonald Observatory.
For detailed description of the spectrographs see \citet{tull95}, the KPNO website\footnote{{\tt http://www.noao.edu/kpno/manuals/echman/}} and \citet{tull98}, respectively.
Exposure times were between 150~s and 1800~s. Typical resolution ($\lambda/\Delta\lambda$) was $\approx 60,000$.

In this work the astrophysical false positives that produce large ($\gg$ 1~\kms) 
RV variation do not require the analysis used for planets \citep{mccullough06}. 
Instead we applied less precise and simpler 
method.
All the spectra had 
the wavelength solutions calculated using the ThAr arc lamp which was observed during 
each observing night. The spectra were cross-correlated 
using \cite{tonry79} method %
with the solar spectrum 
\citep{kurucz84}. For HJS, Mayall and HET telescopes respectively, for each spectrum 
11, 8 and 19 orders were usually used with average lengths of 60\AA, 65\AA~and 59\AA. 
The main sources of the RV errors are: insufficiently accurate wavelength solution, the spectral type mismatch \citep[see e.g. ][]{nidever02},  %
and to a much smaller degree, convective blueshift and gravitational redshift. %

Three stars (J00121\ldots, 
J0813\ldots 
and J1157\ldots) showed evidence for large rotational 
broadening. For these stars telluric lines from H$\alpha$ region were removed 
and projected rotation speed of the star (\vsini) was found by minimizing 
$\chi^2$ between the rotationally broaden solar template and the observed spectrum in 
H$\alpha$ region. This region was chosen because abundance of Hydrogen does not 
change as much as it does for other elements from star to star and it gave 
us many data points. It was repeated for every spectrum of a target star and 
results were averaged to obtain the final \vsini~estimation. For these 3 stars,
we cross correlated their spectra with rotationally broaden solar 
template. 

Some stars exhibit significant RV variations ($\gg$ 1~\kms) and thus they were removed from 
the target list as stellar EBs. For most of them we had 1 or 2 observations from each of  
2 different nights. We assumed our photometric 
ephemerides 
were correct and 
the orbits are circular. For some objects we give better ephemeris in notes for 
Table~\ref{tab:falsepos} and these are the ones used for the calculations. Then, the 
radial velocity ($RV\left(t\right)$) measured at the time ($t$) should follow the relation:
\begin{equation}\label{equ:rveb}
RV\left(t\right)=RV_0-K\cdot\sin\left(\frac{t-T_0}{P}\cdot2\pi\right)
\end{equation}
which allowed us to estimate the values of semiamplitude ($K$) and systemic velocity 
($RV_0$). All of these systems show eclipses so $i\approx90$\arcdeg. We used the
RV semi-amplitude $K$ and the period to determine the mass function ($f\left(m\right)$). 
An approximate mass estimate ($m_{est}$) for the less massive companion follows if we assume its
mass is much smaller than the mass of the primary ($M$): 
\begin{equation}\label{equ:mcrude}
m_{est} = \left(f\left(m\right)\right)^{1/3}M^{2/3}
\end{equation}
To estimate $M$ we used $J-H$ colors from 2MASS, mass-colour relations from \cite{drilling00} 
and assumed the primary is a main sequence star. 
One should note these $m_{est}$ are rough estimates done mainly for choosing targets for follow-up observations.

Table~\ref{tab:ebraw} gives the journal of RV measurements for stars with significant RV 
shift.
shift .
Table~\ref{tab:ebsimple} summarizes results for these binaries. We found no RV data for
these systems in external databases (e.g. ADS and SIMBAD).
For stars with exactly 2 RV measurements, $K$ and 
$RV_0$ were calculated directly and in all other cases we produced a least-squares estimate.
For 2 targets (J00021\ldots and 
J0007\ldots) we were not able to fit a model (Equation 
\ref{equ:rveb}) to the data, because each had a dubious period, considerable measurement uncertainty,
or perhaps due to a non-circular orbit. For J2359\ldots we found negative $K$,
presumably due to an incorrect photometric period or phase.
In both these cases we have estimated a lower limit of $K$ as half of the difference 
between maximum and minimum $RV\left(t\right)$, and associated lower limits of 
$f\left(m\right)$ and $m_{est}$, 
with the caveat that the latter depends on the period. 
There is small possibility that J2359\ldots is a rare example of an EB with components of similar 
effective temperatures, an eccentric orbit, and the smaller star transiting 
the larger one, but the secondary eclipse is not observed due to the inclination.

We have revealed six new double line spectroscopic binaries namely: 
J03482\ldots, 
J0722\ldots,
J0727\ldots, 
J1511\ldots,
J1540\ldots
and J23565\ldots.

\section{Notes on selected stars}
\label{sec:notes}
\paragraph{J00021\ldots}

The period is unreliable. We did not find one which fits well to our photometry and RVs. 
The egress is $\approx$1~h long and only 2 transits were observed (2452912.833 JD and 
2453295.824 JD). A 22~d period was used to estimate $f\left(m\right)$ 
and $m_{est}$.

\paragraph{J00023\ldots}

This object was identified independently by \citet{christian06}. Time of their epoch (taken with weight 3) 
and ET observation were used to refine the ephemeris for transit center: $T_c=2453653.7774+E\cdot2.37554$. 
\citet{christian06} report a companion radius equal to 2.07~R$_J$;
ours is 2.4~R$_J$. \citet{christian06} also report a large
ellipsoidal amplitude. There are two stars of similar brightness 
1\farcs2 apart, thus probably one of them is an EB with twice larger amplitude demonstrating a stellar nature of the ``transiting'' object. 
The centroid shift method applied to the XO data is not sensitive to such small separations.

\paragraph{J0015\ldots}
The brightest object within the XO photometric aperture is \object{2MASS J001524.05+325625.2} ($J = 10.48$). 
We have found significant centroid shift in XO data at $\mathrm{PA} = 339\fdg5$.
The closes star with similar PA is 001523.09+325708.2 ($J = 12.54$, $\mathrm{PA} = 344\fdg4$ and $d = 44\farcs7$), but another possibility is that \object{2MASS J001516.43+325849.4} ($J = 10.67$, $\mathrm{PA} = 326\fdg4$ and $d = 173\farcs2$) causes observed variability.

\paragraph{J040022\ldots}

\begin{figure}[t]
\includegraphics[angle=270,width=.98\columnwidth]{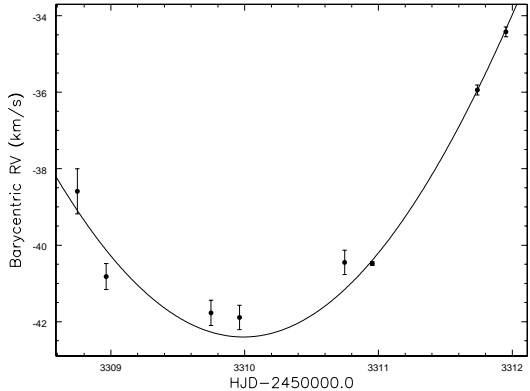}
\figcaption{RV of J040022\ldots. 
Line is defined by Equation~\ref{equ:rveb} with parameters taken from Table~\ref{tab:ebsimple} and ephemeris given in \S \ref{sec:notes}.\label{fig:rvplot1}}
\end{figure}

The XO observed only 2 partial transits of this object, separated by 34~d. 
Both ingress and egress last $\approx$3 h. We observed a local minimum of RV with the Mayall 
telescope one season later than the latest XO photometry (see Figure~\ref{fig:rvplot1}). 
The ephemeris which fits all of our data, $T_c=2452964.12+E\cdot17.08$, was used 
for calculations %
in Table~\ref{tab:ebsimple}. The \bv~$=1.0$, the eclipse is \U-shaped with long ingress 
and egress. The best solution we have found it is a K giant orbited by a dwarf.

\paragraph{J1157\ldots}
This candidate was identified independently by \citet{kane08}.
ET observations combined with XO survey data give the ephemeris: $T_c=2453436.1140+E\cdot2.45379$. 
Our RV measurements show its eclipses are not caused by a planet because it exhibits
30~\kms~shift in 1~d.

\paragraph{J1515\ldots}
For this object one can see out of transit variations in XO data. Only two transits were observed and periods $16.85488/i$~d ($i$ is an integer) are consistent with photometric data. Four RV measurements constrained the period. Best fitting ephemeris is $T_c=2453866.0476+E\cdot5.61829$ and it was used to obtain values presented in Table~\ref{tab:ebsimple}.

\paragraph{J1516\ldots}

\begin{figure}[t]
\plotone{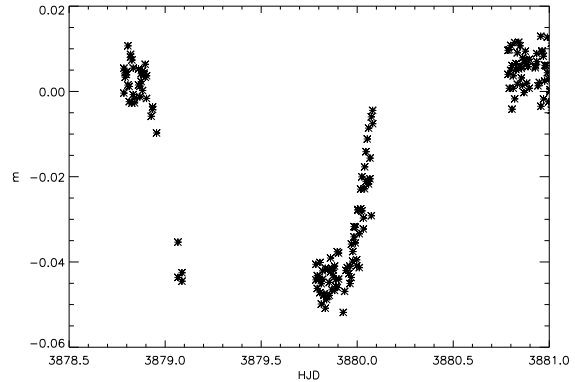}
\figcaption{XO light curve for 3 nights of J1516\ldots. Data were 
calibrated using SysRem algorithm. Size of points does not indicate errors. 0~mag 
corresponds to mean brightness.\label{fig:egress}}
\end{figure}

We have observed only a few transit events for this star. The best observation showed egress 
beginning 2453878.95~JD and ending 2453880.10~JD. Thus egress lasts $\approx$3.5~h and the 
whole transit is $\approx$27.5~h (see Figure~\ref{fig:egress}) so we interpret it as a long period system. 
The shortest time difference between transits observed by XO is $\approx 26$~d, 
{\it i.e.} three times longer than value found using BLS.
The fact that eclipses are observed implies inclination close to $90 \degr$. If we also assume mass of the primary star $1 M_{\odot}$ than the crude estimate \cite[Eq. 2][]{mccullough06} of the stellar radius is $6R_{\odot}$.
I depends on period assumed: a period of 50~d gives half the former radius. 
We assign this candidate to be a giant orbited by a main sequence star.

\section{Conclusions}
\label{sec:conc}
This paper presents list of \totalno stars with light curves 
that mimic planetary transits but the follow-up investigation 
demonstrated they are instead astrophysical false positives:
grazing-incidence orbits of EBs, eclipses of a large star by an M dwarf, 
or eclipsing systems whose light has been diluted by a nearby bright star. 
Fifteen light curves were classified as \V-shape and fifteen as \U-shape.
The list can be helpful for others searching for transiting planets,
and also for eclipsing binaries, with low-mass stellar companions.

\begin{figure}[t]
\plotone{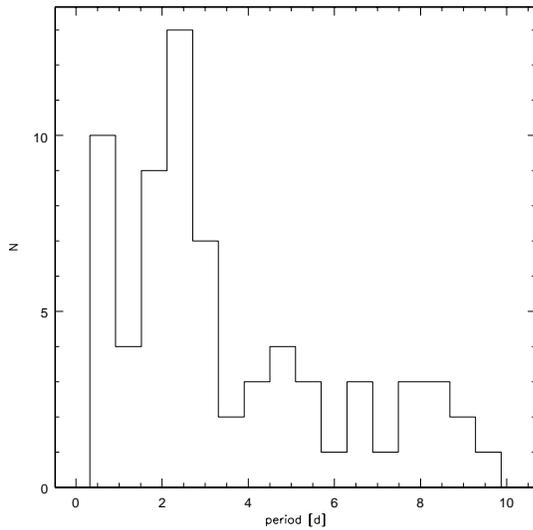}\figcaption{Histogram of periods presented in Table~\ref{tab:falsepos}.\label{fig:periods}}
\end{figure}

Most of the stars in Table~\ref{tab:falsepos} are located in the strips centered 
at 0 and 4 hours RA. Those strips were the first to be observed. 
With greater experience, our improved
algorithms \citep{mccullough07} produced a smaller number of astrophysical 
false positives from other strips. Transit depths found by the BLS algorithm range from 4~mmag to 69~mmag.
In Figure \ref{fig:periods} we present histogram of periods presented in Table~\ref{tab:falsepos}. Most of astrophysical false positives have periods shorter than 3.5~d. Obviously, probability of detecting short lasting events like transits is lower for longer period objects.

RVs presented in this paper have 
typicall uncertainties of 400 m/s.
For \onelineno single-line spectroscopic binaries we have 
estimated mass function. 
Six new double-line spectroscopic binaries were revealed. 
Additional remarks for some objects were given.

The most interesting objects for follow-up observations, which aim is to find low-mass secondaries of EBs, are systems with the smallest value of mass function  given in Table~\ref{tab:ebsimple}. 
Among the stars without RV measurements the ones with the smallest transit depths and \U-shaped eclipses should be the most promising.

\acknowledgments 
R. P. is grateful to the AURA for the Summer Student Program at STScI as well as 
I. Soszy{\'n}ski for comments on manuscript.
This publication makes use of catalogs VizieR and SIMBAD operated 
at CDS, Strasburg, France, and data products from the 2MASS, 
which is a joint project of the University of Massachusetts 
and the IPAC/Caltech, funded by the NASA and NSF.

\begin{deluxetable}{crrrrrrrcc}

\tabletypesize{\small} 
\tablecaption{Astrophysical false positives found by X0 \label{tab:falsepos}}
\tablewidth{0pt}
\tablecolumns{9}
\tablehead{\colhead{} & \colhead{$m_{XO}$} & \colhead{} & \colhead{$\delta m$} & \colhead{$\delta m_{FL75}$} & \colhead{$t_{dur}$} & \colhead{P} & \colhead{$T_0$} & \colhead{} \\
\colhead{2MASS} & \colhead{(mag)} & \colhead{FL75} & \colhead{(mag)} & \colhead{(mag)}&\colhead{(h)} & \colhead{(d)} & \colhead{JD-2450000.0} & \colhead{Flags} \\}
 
\startdata 
{\it \object{J000210.00+474803.6}} & 10.98 & 0.93 & 0.011 & 0.012 & 3.68 & 9.57487 & 3295.817 & SB1 \\
{\it \object{J000233.06+331517.3}} & 10.96 & 0.47 & 0.022 & 0.047 & 1.82 & 2.37552 & 3653.765 & \V \\ 
\object{J000629.13+355242.4} &  9.84 & 0.08 & 0.004 & 0.052 & 1.46 & 0.897843 & 3319.745 & \nodata \\  
\object{J000744.08+402835.4}\tablenotemark{a} & 10.17 & 0.93 & 0.015 & 0.016 & 2.19 & 2.84981 & 2948.804 & SB1 \\
\object{J000857.97+025642.0}\tablenotemark{b} & 10.12 & 1.00 & 0.011 & 0.011 & 3.17 & 4.72277 & 3653.778 & SB1 \\
\hline
\object{J001215.64+335112.1} & 11.33 & 0.96 & 0.011 & 0.012 & 3.14 & 6.55222 & 3295.906 & SB1 \\
\object{J001222.06+142422.6}\tablenotemark{c} & 12.08 & 0.96 & 0.050 & 0.053 & 2.06 & 1.07386 & 3287.018 & \V \\
\object{J001446.58+301646.2}\tablenotemark{d} &  9.39 & 0.86 & 0.042 & 0.048 & 1.87 & 2.16132 & 3318.841 & \U \\
{\it \object{J001523.09+325708.2}} & 11.56 & 0.15 & 0.036 & 0.261 & 2.58 & 1.79057 & 3320.819 & \ps, \V \\
\object{J033955.56+452329.5} &  9.43 & 0.95 & 0.017 & 0.018 & 3.37 & 8.77347 & 3322.084 & SB1 \\
\hline
\object{J034610.90+283212.5} & 10.57 & 0.22 & 0.017 & 0.080 & 2.81 & 4.87377 & 3300.980 & SB1, \V \\
\object{J034829.26+052719.5} & 11.23 & 0.94 & 0.021 & 0.022 & 3.50 & 4.56329 & 3328.900 & SB2\\
\object{J034854.48+420725.0}\tablenotemark{e} & 10.41 & 0.97 & 0.031 & 0.032 & 3.40 & 3.22290 & 3312.044 & \V \\
\object{J035046.19+454253.9} & 11.49 & 0.06 & 0.015 & 0.253 & 3.05 & 1.67403 & 3369.903 & \V \\
\object{J035129.86+460956.8}\tablenotemark{f} & 11.83 & 0.61 & 0.069 & 0.115 & 3.64 & 7.59071 & 3350.973 & \U \\ 
\hline
\object{J035215.14+545101.3}\tablenotemark{g} & 12.09 & 0.19 & 0.021 & 0.119 & 3.14 & 4.67552 & 3351.941 & \nodata \\
\object{J035308.94+053633.1} & 10.98 & 1.00 & 0.015 & 0.015 & 3.29 & 6.86153 & 3356.835 & SB1, \U \\
\object{J035403.37+150830.3} &  9.04 & 0.98 & 0.024 & 0.025 & 2.07 & 7.19632 & 3657.953 & SB1 \\
\object{J035635.76+274551.0} & 11.69 & 0.51 & 0.031 & 0.061 & 2.31 & 2.67151 & 3375.746 & \ps, \V \\
\object{J035747.56+341415.4} & 11.28 & 0.06 & 0.022 & 0.413 & 2.89 & 1.67034 & 3376-788 & \V \\
\hline
\object{J035839.54+071617.8}\tablenotemark{h} & 11.26 & 0.01 & 0.016 & \nodata & 3.18 & 2.20751 & 3323.031 & \nodata \\ 
\object{J035931.11+011806.1}\tablenotemark{i} & 11.99 & 0.99 & 0.010 & 0.010 & 1.28 & 0.782130 & 3654.135 & \nodata \\
\object{J035954.57+424555.7}\tablenotemark{j} & 10.82 & 0.48 & 0.023 & 0.047 & 3.77 & 7.86040 & 3652.139 & \nodata \\ 
\object{J040016.44+540120.3} & 10.62 & 0.92 & 0.039 & 0.042 & 2.24 & 1.94803 & 3322.124 & \ps, \V \\ 
\object{J040025.10+022526.6}\tablenotemark{k} & 11.34 & 0.13 & 0.016 & 0.127 & 3.33 & 5.77968 & 3376.784 & \nodata \\
\hline 
{\it \object{J040022.90+090533.6}} & 11.32 & 1.00 & 0.028 & 0.028 & 3.29 & 8.51064 & 2997.898 & SB1, \U \\ 
\object{J040052.03+532245.9}\tablenotemark{l} &  9.42 & 0.44 & 0.019 & 0.044 & 3.26 & 2.12368 & 3342.858 & \ps, \V \\
\object{J040117.74+492843.0} & 10.72 & 0.80 & 0.037 & 0.047 & 3.34 & 8.70322 & 3375.779 & \ps \\
\object{J040425.98+480532.2} & 12.00 & 0.10 & 0.018 & 0.199 & 2.42 & 1.57639 & 3321.955 & \nodata \\
\object{J040559.26+512758.7} & 11.47 & 0.32 & 0.012 & 0.039 & 1.85 & 0.622076 & 3658.103 & \V \\
\hline
\object{J040905.12+180323.7} & 12.16 & 0.28 & 0.016 & 0.059 & 2.54 & 1.55560 & 3348.984 & \V \\
\object{J040919.39+485819.3} & 12.25 & 0.14 & 0.047 & 0.394 & 2.20 & 1.62185 & 3293.096 & \nodata \\
\object{J041114.76+145420.9} & 11.77 & 0.23 & 0.015 & 0.068 & 3.19 & 2.21474 & 3350.764 & \V \\ 
\object{J041144.50+311726.1} & 10.99 & 0.55 & 0.044 & 0.081 & 2.91 & 3.03067 & 3349.022 & \U \\
\object{J041314.80+530431.9} & 10.74 & 0.12 & 0.011 & 0.089 & 3.22 & 4.19252 & 3369.894 & \nodata \\
\hline
\object{J041326.43+443227.7}\tablenotemark{m} & 10.63 & 0.92 & 0.013 & 0.015 & 3.30 & 2.64145 & 3655.061 & \U \\
\object{J041807.44+590552.7}\tablenotemark{n} & 9.80 & 0.87 & 0.030 & 0.035 & 1.46 & 3.04952 & 3658.079 & \nodata \\	
\object{J072222.81+255627.1} & 10.81 & 0.99 & 0.014 & 0.014 & 1.49 & 2.22311 & 4153.780 & SB2 \\ 
\object{J072706.06+311708.9}\tablenotemark{o} &  9.85 & 0.75 & 0.016 & 0.022 & 3.57 & 4.12712 & 4152.766 & SB2, \U \\ 
\object{J073625.33+614626.3} & 11.95 & 0.04 & 0.007 & 0.184 & 2.23 & 0.775711 & 3409.817 & \nodata \\
\hline
\object{J074506.62+444650.6}\tablenotemark{p} &  9.83 & 0.03 & 0.007 & 0.269 & 1.82 & 0.860526 & 3398.780 & \V \\
\object{J074943.04+005230.3} & 10.66 & 0.08 & 0.007 & 0.081 & 2.92 & 1.26759 & 3417.806 & \nodata \\
\object{J080856.73+214452.8} & 10.24 & 0.96 & 0.025 & 0.026 & 2.99 & 5.19481 & 3412.027 & SB1, \U \\
\object{J081137.72+212013.8}\tablenotemark{q} & 10.85 & 0.05 & 0.017 & 0.376 & 2.34 & 2.44248 & 3419.821 & \V \\
\object{J081338.69+372352.4}\tablenotemark{r} & 11.95 & 0.50 & 0.009 & 0.017 & 1.82 & 2.71400 & 3436.794 & SB1, \V \\
\hline
\object{J114634.97+544538.8} & 11.96 & 0.83 & 0.017 & 0.021 & 2.05 & 3.56608 & 3491.831 & \V \\
{\it \object{J115718.68+261906.1}} & 11.10 & 0.89 & 0.013 & 0.015 & 2.59 & 2.45387 & 3436.114 & SB1, \U \\ 
\object{J151110.13+385703.1}\tablenotemark{s} &  8.96 & 1.00 & 0.021 & 0.021 & 1.88 & 1.50525 & 4206.083 & SB2, \V \\
{\it \object{J151559.79+503139.4}} & 10.77 & 0.99 & 0.018 & 0.018 & 3.23 & 8.42744 & 3866.048 & SB1 \\
{\it \object{J151623.71+090139.2}} & 10.27 & 0.97 & 0.045 & 0.047 & 4.93 & 8.55725 & 3879.850 & \U \\
\hline
\object{J151843.24+533338.8SW}\tablenotemark{t} &  9.66 & 1.00 & 0.025 & 0.025 & 3.28 & 3.79276 & 4221.107 & SB1, \U \\
\object{J152327.59+023329.6} &  9.65 & 0.02 & 0.007 & 0.641 & 1.78 & 0.883986 & 4234.964 & \nodata \\
\object{J153248.92+070945.1} & 11.94 & 0.12 & 0.018 & 0.166 & 1.24 & 0.807637 & 4200.965 & \nodata \\
\object{J154046.82+621339.6} & 12.01 & 0.98 & 0.020 & 0.020 & 2.08 & 3.09962 & 3884.971 & SB2 \\
\object{J155618.13+074537.6} & 11.97 & 0.05 & 0.010 & 0.236 & 3.21 & 2.39173 & 3506.011 & \nodata \\ 
\hline
\object{J234031.68+474558.6} & 11.01 & 0.14 & 0.006 & 0.043 & 2.05 & 0.713328 & 3283.938 & \nodata \\
\object{J234512.08+343544.5}\tablenotemark{u} & 12.45 & 0.85 & 0.037 & 0.043 & 2.44 & 6.35001 & 3293.848 & \nodata \\
\object{J234822.41+185717.4}\tablenotemark{v} & 12.38 & 0.75 & 0.067 & 0.091 & 3.02 & 5.24164 & 3652.862 & \V \\
\object{J234837.19+181348.7} &  9.79 & 0.97 & 0.041 & 0.042 & 3.75 & 7.81983 & 3665.835 & \ps, \U \\
\object{J234959.04+311203.5} & 11.88 & 0.14 & 0.017 & 0.130 & 2.70 & 1.48091 & 3241.953 & \nodata \\
\hline
\object{J235035.06+294350.3} & 12.04 & 0.85 & 0.025 & 0.030 & 3.38 & 2.93582 & 3650.769 & \ps, \V\\
\object{J235048.78+440127.2}\tablenotemark{w} & 11.61 & 0.92 & 0.022 & 0.024 & 1.33 & 0.815248 & 3319.738 & \V \\ 
\object{J235104.27+251629.3}\tablenotemark{x} & 12.40 & 0.53 & 0.030 & 0.058 & 2.43 & 2.53075 & 2912.854 & \U \\
\object{J235219.65+434323.4} & 11.82 & 0.15 & 0.018 & 0.127 & 1.65 & 0.659657 & 3295.857 & \nodata \\
\object{J235227.06+395515.1} & 11.87 & 0.02 & 0.008 & 0.391 & 2.94 & 1.53116 & 3321.854 & \nodata \\
\hline
\object{J235602.52+415451.5} & 11.75 & 0.71 & 0.033 & 0.046 & 2.70 & 2.16572 & 3652.945 & \ps, \U \\
\object{J235613.98+402648.3}\tablenotemark{y} & 12.20 & 0.55 & 0.022 & 0.040 & 2.54 & 4.41618 & 3318.862 & \ps \\
\object{J235650.15+160754.7} & 12.15 & 0.88 & 0.039 & 0.045 & 2.60 & 1.59084 & 3651.045 & SB2, \U \\
\object{J235929.73+444031.2} & 10.61 & 0.97 & 0.032 & 0.033 & 3.27 & 5.68117 & 3320.788 & SB1, \U \\
\enddata

\tablecomments{Astrophysical false positives found by the X0 project. 
Additional comments for stars marked by italics can be 
found in Section \ref{sec:notes}. The flags are: 
(\ps) --- evidence for different primary and secondary minima if the light curve is folded with double period, 
(SB1) --- spectroscopic single line binary, 
(SB2) --- spectroscopic double line binary, 
(\U) --- transit is \U-shaped, 
(\V) --- transit is \V-shaped. 
}

\tablenotetext{a}{The Period is $39.897/i$ where $i=1,2,3,6,7,14$ and $i=14$ seems most probable.}
\tablenotetext{b}{Only 2 transits 344.02~d apart were observed with different depths.}
\tablenotetext{c}{The XO amplitude is 0.07~mag and it is too big for a planetary transit.}
\tablenotetext{d}{The ET measured the $R$ amplitude 0.065~mag which is too big for a planetary transit.}
\tablenotetext{e}{The $R$ amplitude $\geq$0.07~mag and the transit lasts at least 6~h, so it is too deep and too long for a planetary transit.}
\tablenotetext{f}{ET observations showed the $R$ amplitude $\geq$0.11~mag and were used to find better ephemeris: $T_c=2453550.9416+E\cdot7.58945$.}
\tablenotetext{g}{The ET observations showed it is an EB with different depths and half period. The ephemeris is: $T_c=2453351.9538+E\cdot2.33781$.}
\tablenotetext{h}{
$\delta m_{FL75}$ 
is smaller than $\delta m$. It shows our estimation of photometric aperture is not working well in this case. ET found R amplitude $\geq 0.80$~mag.}
\tablenotetext{i}{Secondary eclipses were observed by the XO.}
\tablenotetext{j}{The ET found $R$ amplitude $\geq$0.058~mag. There is a bright stars nearby. The ingress and the egress lasts $\approx$2~h each. Most probable periods are 15.7191~d and 7.8595~d.}
\tablenotetext{k}{If double period is used than out of transit variation is seen on the XO data. There is brighter nearby star \object{2MASS J040026.26+022555.7} and centroid shift was found at $\mathrm{PA}=239\fdg0$. The star 040025.10+022526.6 is situated $d = 34\farcs7$ apart at $\mathrm{PA}=237\fdg0$.} 
\tablenotetext{l}{\citet{fabricius02} found this is a visual binary and gave it designation \object{TDSC 8509}. The brighter component is \object{2MASS J040049.67+532237.3}. Centroid shift was found in XO data at $\mathrm{PA}=17\fdg2$. We revealed \object{TDSC 8509 B} (2MASS J040052.03+532245.9) to be a binary system itself. It is located $d=36\farcs4$ apart at $\mathrm{PA}=13\fdg7$.} 
\tablenotetext{m}{If we assume this star to have Jupiter size planet than given period gives us estimation of transit length of $\approx$1.9~h and length measured on the XO data is too long. ET show even longer event lasting $\approx$4~h.}
\tablenotetext{n}{The XO observed only two ingresses which are separated by 365.91~d and last $\approx$1.5~h. This object has definitely a very deep eclipses and probably period of few times longer than 3.05~d.}
\tablenotetext{o}{This is known double star \object{TDSC 19502} \cite{fabricius02}.} 
\tablenotetext{p}{The ET found filter-dependent amplitudes: 0.47~mag in $B$, 0.40~mag in $V$, 0.35~mag in $I$ and 0.40~mag in $R$. Primary and secondary minima are observed.}
\tablenotetext{q}{Better ephemeris was found using the ET observations: $T_c=2453419.8112+E\cdot2.44269$.}
\tablenotetext{r}{Better ephemeris was found using the ET observations: $T_c=2453436.7858+E\cdot2.71387$.}
\tablenotetext{s}{Shows different depth transits when phased with four times longer period. Out of transit variation is observed.}
\tablenotetext{t}{This is known double star. It has one designation in 2MASS thus FL75 is 1.0, but should be $\approx$0.5. We are designating components: 151843.24+533338.8SW (CCDM15187+5334A) and 151843.24+533338.8NE (CCDM15187+5334B). See \cite{dommanget02} for more information.}
\tablenotetext{u}{The ET observations show the transit at least 0.05~mag deep so companion is thus at least 2R$_J$ what is too big for a planet.}
\tablenotetext{v}{The XO amplitude is 0.09~mag.}
\tablenotetext{w}{The ET found filter-dependent amplitudes: 0.02~mag in $B$, 0.03~mag in $V$ and 0.04~mag in $R$.}
\tablenotetext{x}{The ET found that $R$ amplitude is 0.07~mag.}
\tablenotetext{y}{The Period is dubious (may be even 5 times longer) because only 1 ingress and 2 egresses were observed. The XO amplitude is 0.03~mag. The ET found $R$ and $V$ amplitudes are 0.04~mag and transit duration is at least 3~h.}
\end{deluxetable}

\begin{deluxetable}{lrlrr}
\tabletypesize{\small}
\tablecaption{RV measurements for EBs \label{tab:ebraw}}
\tablewidth{0pt}
\tablecolumns{5}
\tablehead{\colhead{2MASS} & \colhead{$t$} & \colhead{} & \colhead{$RV\left(t\right)$} & \colhead{$\sigma_{RV}$}\\
\colhead{designation} & \colhead{(HJD-2450000.0)} & \colhead{Telescope} & \colhead{(\kms)} & \colhead{(\kms)}}

\startdata
J00021\ldots & 3308.6305 & Mayall & -1.18 & 0.14 \\
J00021\ldots & 3309.6245 & Mayall & -1.65 & 0.33 \\
J00021\ldots & 3310.6160 & Mayall & -3.92 & 0.12 \\
J00021\ldots & 3310.8680 & Mayall & -4.84 & 0.13 \\
J00021\ldots & 3311.6177 & Mayall & -6.10 & 0.22 \\
J00021\ldots & 3954.7807 & HET & -27.28 & 0.12 \\
J00021\ldots & 3954.7845 & HET & -27.54 & 0.27 \\
J00021\ldots & 3958.8006 & HET & -22.57 & 0.06 \\
J00021\ldots & 3958.8044 & HET & -22.58 & 0.08 \\
J00021\ldots & 3965.7672 & HET & -1.90 & 0.21 \\
J00021\ldots & 3965.7709 & HET & -1.88 & 0.23 \\
J0007\ldots  & 3308.6537 & Mayall & -23.56 & 0.30 \\
J0007\ldots  & 3308.8803 & Mayall & -25.83 & 0.15 \\
J0007\ldots  & 3309.6477 & Mayall & -29.88 & 0.06 \\
J0007\ldots  & 3309.8735 & Mayall & -31.72 & 0.17 \\
J0007\ldots  & 4321.8032 & HET & -4.81 & 0.17 \\
J0007\ldots  & 4322.7790 & HET & -6.47 & 0.10 \\
J0008\ldots  & 3308.7077 & Mayall & 8.03 & 0.21 \\
J0008\ldots  & 3308.8137 & Mayall & 5.21 & 0.20 \\
J0008\ldots  & 3309.7035 & Mayall & -6.26 & 0.31 \\
J0008\ldots  & 3309.8146 & Mayall & -8.40 & 0.15 \\
J00121\ldots & 4309.8301 & HET & -0.7 & 5.6 \\
J00121\ldots & 4318.8004 & HET & -33.9 & 2.8 \\ 
J0339\ldots  & 3997.8284 & HET & -25.0 & 1.7 \\
J0339\ldots  & 4003.8122 & HET & -8.7 & 2.1 \\
J0346\ldots & 3310.7882 & Mayall & -1.11 & 0.77 \\
J0346\ldots & 3311.8722 & Mayall & -42.45 & 0.17 \\
J0353\ldots  & 3308.7732 & Mayall & -29.86 & 0.53 \\
J0353\ldots  & 3308.9357 & Mayall & -32.01 & 0.27 \\
J0353\ldots  & 3309.7733 & Mayall & -42.43 & 0.37 \\
J0353\ldots  & 3309.9330 & Mayall & -43.05 & 0.24 \\
J0354\ldots  & 3997.8496 & HET & -29.99 & 0.07 \\
J0354\ldots  & 4001.8444 & HET & 6.55 & 0.17 \\
J040022\ldots  & 3308.7492 & Mayall & -38.59 & 0.59 \\
J040022\ldots  & 3308.9650 & Mayall & -40.82 & 0.34 \\
J040022\ldots  & 3309.7484 & Mayall & -41.77 & 0.33 \\
J040022\ldots  & 3309.9615 & Mayall & -41.89 & 0.32 \\
J040022\ldots  & 3310.7467 & Mayall & -40.45 & 0.32 \\
J040022\ldots  & 3310.9537 & Mayall & -40.48 & 0.05 \\
J040022\ldots  & 3311.7390 & Mayall & -35.94 & 0.13 \\
J040022\ldots  & 3311.9523 & Mayall & -34.42 & 0.13 \\
J0808\ldots  & 4141.8338 & HJS & 48.04 & 0.32 \\
J0808\ldots  & 4142.8604 & HJS & 61.19 & 0.51 \\
J0808\ldots  & 4143.8531 & HJS & 56.84 & 0.83 \\
J0813\ldots & 4140.9303 & HJS & -24.8 & 1.7 \\
J0813\ldots & 4178.7487 & HET & -39.6 & 1.1 \\ 
J0813\ldots & 4179.7464 & HET & 21.9 & 1.5 \\ 
J1157\ldots  & 4225.7663 & HET & 23.32 & 0.78 \\
J1157\ldots  & 4226.7732 & HET & -8.67 & 0.45 \\
J1515\ldots  & 4141.0176 & HJS & 20.42 & 0.26 \\ 
J1515\ldots  & 4141.9549 & HJS & -30.58 & 0.30 \\
J1515\ldots  & 4142.9524 & HJS & -42.41 & 0.88 \\
J1515\ldots  & 4144.9652 & HJS & 45.15 & 0.35 \\
J1518\ldots & 4139.0004 & HJS & -69.89 & 0.27 \\
J1518\ldots & 4140.9836 & HJS & -28.24 & 0.17 \\
J1518\ldots & 4141.9675 & HJS & -68.20 & 0.22 \\
J1518\ldots & 4142.9678 & HJS & -64.36 & 0.60 \\
J2359\ldots  & 3310.6391 & Mayall & -19.07 & 0.42 \\
J2359\ldots  & 3311.7990 & Mayall & -26.36 & 0.50 \\
\enddata

\tablecomments{RV measurements for EBs. See \S \ref{sec:rv} for description of weights and data reduction.}
\end{deluxetable}

\begin{deluxetable}{lrrrrrr}
\tabletypesize{\small}
\tablecaption{RV parameters for EBs \label{tab:ebsimple}}
\tablewidth{0pt}
\tablecolumns{7}
\tablehead{\colhead{2MASS} & \colhead{\vsini} & \colhead{$K$} & \colhead{$RV_0$} & \colhead{$f\left(m\right)$} & \colhead{$M$} & \colhead{$m_{est}$} \\
\colhead{designation} & \colhead{(\kms)} & \colhead{(\kms)} & \colhead{(\kms)} & \colhead{($M_\sun$)} & \colhead{($M_\sun$)} & \colhead{($M_\sun$)} }

\startdata 
{\it J00021\ldots} & \nodata & $\ge$13.10 & \nodata & $\ge$0.0051 & 1.3 & $\ge$0.17 \\ 
J0007\ldots & \nodata & $\ge$13.12 & \nodata & $\ge$0.00067 & 1.4 & $\ge$0.087\\ 
J0008\ldots & \nodata & 12.39 & 2.68 & 0.00093 & 1.5 & 0.13 \\ 
J00121\ldots & 56.5$\pm$0.5 & 20.31 & -21.01 & 0.0057 & 1.5 & 0.23 \\ 
J0339\ldots & \nodata & 15.20 & -23.16 & 0.0032 & 1.5 & 0.19\\ 
J0346\ldots & \nodata & 45.05 & 2.39 & 0.046 & 1.0 & 0.36 \\ 
J0353\ldots & \nodata & 15.12 & -30.22 & 0.0025 & 1.3 & 0.16 \\
J0354\ldots & \nodata & 18.57 & -11.56 & 0.0048 & 1.4 & 0.21 \\
{\it J040022\ldots} & \nodata & 32.29 & -10.11 & 0.060 & 0.9 & 0.36 \\ 
\nodata & \nodata & \nodata & \nodata & \nodata & 1.0 & 0.39 \\ 
J0808\ldots & \nodata & 12.59 & 49.30 & 0.0011 & 1.6 & 0.14 \\ 
J0813\ldots & 39.9$\pm$1.0 & 37.31 & -15.22 & 0.015 & 1.4 & 0.31 \\ 
{\it J1157\ldots} & 40.1$\pm$2.0 & 16.66 & 7.47 & 0.0012 & 1.5 & 0.14 \\ 
{\it J1515\ldots} & \nodata & 52.28 & 2.49 & 0.083 & 1.5 & 0.57 \\ 
J1518\ldots & \nodata & 27.56 & -47.70 & 0.0082 & 1.0 & 0.20 \\ 
J2359\ldots & \nodata & -15.22 & -33.90 & $\ge$0.000029 & 1.4 & $\ge$0.031 \\
\enddata

\tablecomments{RV parameters for EBs. See \S \ref{sec:rv} for description of analysis and \S \ref{sec:notes} for comments concerning stars marked by italics. For J040022\ldots two estimates are given. For first one we have assumed its luminosity class is V, for second -- class III.}
\end{deluxetable}

\end{document}